\documentstyle[12pt,epsf]{article}
\catcode`\@=11 \@addtoreset{equation}{section}  
\renewcommand{\theequation}                     
         {\arabic{section}.\arabic{equation}}   
\title{ Underlying Field Structure of Matter }
\author{P. Leifer}
\date{I.T.L. Ltd., POB 211 \\
Or-Yehuda 60251, Israel} 
\begin{document}
\maketitle
\begin{abstract}
Special relativity (SR), general relativity (GR), and quantum theory
(QT) were developed on the ground of classical mechanics.
There is, however, the deep internal incompatibily 
between these theories. On one hand SR and GR are 
intrinsically deterministic in spacetime since they deal
with pure classical notions (e.g. material point, events,
etc.) On the other hand in QT one uses deterministic equations 
in a state space (e.g. in Hilbert space) but 
loses a possibility of deterministic description in spacetime.
Attempts to avoid contradictions between these theories
evoke a new approach to their intrinsic unification. The 
unification may be achieved only on the base of 
the quantum notions which are deeper than classical ones. 

Herein I will discuss application of such approach to 
quantum dynamical model of a fundamental field as a presumably 
robust model of the unified interaction. At this 
stage only general picture of that model can be described. 
\end{abstract}
\vskip .2cm
\section{Introduction}
The question is: what the quantum level should be used for
the desirable unification? Attempts based on established
spacetime structure where present
quantum particles with pointwise interaction lead to big 
difficulties. Probably ``elementary'' particles level is acceptable
since ``every elementary particle consists of all other elementary
particles'' and, therefore, presumably does exist some unified
field entity (let us say ``fundamental field'' (FF)) which excitations 
represent observable particles. This FF, I think, paves the 
natural way to desirable unification. One known 
example of such theory of nonlinear fundamental spinor field
was proposed by Heisenberg \cite{Heisen}. The differences between
this approach and our are as follow:
 
1. We avoid to endow our FF by some spacetime properties on
   the same ground as Shapere and Wilczek introduced ``unlocated
   shapes'' of deformable body \cite{SW}. In our case we will
   deal with deformable quantum state (configuration) of FF.
   The reason is based on the assumption that self-dynamics of 
   FF is hidden, does not depend on position in Universe, and,
   therefore, does not belong in literal sense  
   ``a priori'' to spacetime manifold. Maybe better to say that 
   spacetime coordinates indeed do not exist a priori.  
   Only interaction with ``measuring device'' leads to 
   arithmetization (coordinatization) of spacetime structure. 

2. The mutual transformations of one internal degree of freedom 
   to another (isospin to spin, for example) give a possibility 
   to build theory of scalar FF. This is a relatively new 
   effect of nonlinear 
   field theory \cite{Ait} was not well known when Heisenberg
   tried to build his theory.
   In that sense our approach is close to Skyrme assumption about
   fundamental role of boson field \cite{Skyrme1,Skyrme2}. Our
   target is to establish transformation (map) of internal degrees
   of freedom into external (spacetime) ones. In other words
   such transformation should map spin, charge, etc., into energy-
   momentum of observable quantum particles. 
   
3. Gauge fields are not induced by the local (in spacetime) 
   transformations of amplitudes but arise due to introduction
   of local coordinates in 
   projective state space which hereby ignore global general phase.
   They are not independent from "matter fields", have the same 
   origin, and related to the affine connection of the
   Hilbert projective state space $CP(N)$. This affine connection 
   like in GR ``is itself constructed from first derivatives of metric
   tensor'' \cite{Weinberg1}. In our case the Fubini-Study metric 
   of $CP(N)$ and its connection play major role in state-dependent
   gauge theory \cite{Le1,Le2,Le3} 

4. Local (in projective state space) invariants of the isotropy
   group of coherent state of FF will be identified with electric charge,
   strange, etc. The principle difference between our model and
   previously proposed schemes is that we are not dealing with 
   multiplete of ``elementary'' particles but with {\bf multiplete of
   elementary invariants (charges) of the internal FF symmetry 
   groups which now will be treated not phenomenological 
   but fundamental}. Under some experimental condition components 
   of such multiplete may be identified with an ``elementary'' particle.

5. Variational principle I will formulate in tangent fiber bundle
   over $CP(N)$, not in spacetime itself. 

Some explanations are desirable for clarification big
differences between out approach and commonly used ideology. 
 
The most fundamental notion of SR and GR is {\it event}
which is a {\it quantum transition} from the quantum point of view. 
The last has objective sense and clear geometric interpretation
in complex projective state space $CP(N)$. Such approach has 
been based on geometry of the pure quantum states space 
of vacuum excitations of unified field and was called 
``superrelativity''.  Mainly it is a new nonlinear 
version of quantum field theory where quantum
state represented by local field variables are primary and spacetime 
position is secondary. Therefore all physically important
notions like energy, action and even relative spacetime coordinates
should be expressed in terms of local coordinates of $CP(N)$.
Hereby the structure of the {\bf physical space} will be 
completely different in comparison with previous physical
theories. Namely, projective Hilbert space takes place of the 
base manifold, and spacetime coordinates are merely very 
special coordinates in the fiber of tangent fiber bundle over 
this manifold.      
 
The main physical content of superrelativity is a possibility
to compensate (create) unified physical field to (from) vacuum state
of FF by the choice of local frame in Hilbert projective state space
$CP(N)$ rather than by the choice of reference frame in spacetime. 
This local frame formalizes, in fact, experimental environment
influence, deforming the vacuum state. The deformed states
properties are defined by the coset structure 
$G/H=SU(N+1)/S[U(1) \times U(N)]$ of the vacuum state excitations.

It is often emphasized that in quantum theory the
notion of forces has restricted meaning and that the 
energy of  interaction is much more adequate notion. But the 
operator of energy (Hamiltonian) one must usually
build by  the method of classical analogy. This is essentially
ambiguous procedure since quantum energy is the
function of canonical dynamical variables which do not
commute in the general case. 
Big efforts were spent in order to find the
method of the natural transformation of classical 
dynamical variables in corresponding quantum observables
and vice versa. 
I will chose a different approach: {\bf to introduce
inherently quantum classification of quantum motions and
adequate dynamical variables generating different kinds
of such motions} \cite{Le1}.
In order to do it we ought to recognize the 
non-formal difference between quantum 
and classical dynamical variables.
The different character of dynamical variables in 
quantum and classical physics associated with 
different character of internal and external degrees of freedom.
Spacetime in classical physics is the fundamental
universal structure since any classical system
can be represented as a set of the material points.
That is only dynamical variables corresponding
spacetime degrees of freedom are fundamental and
the dynamical variables of any mechanical system are 
reducible to energy-momentum and to angular moment
of subsystems. Therefore classical phase space or 
configuration spaces are merely auxiliary entities. 
In this sense the note of Einstein \cite{Einstein1} about 
the more  natural character the expression of interaction 
in spacetime than in the configuration space of 
Schr\"odinger's \cite{Sch1} quantum mechanics, 
is quite adequate. But at the more deep
quantum level (e.g. elementary particles) one has, 
rather, the opposite situation. 
It means that we have not, in fact, material points 
which were used in Schr\"odinger's geometric model
of configuration space. They presumably are some 
soliton-like solutions of non-linear wave 
equations which must be obtained from the first 
principles concerning internal structure
of quantum ``elementary'' particles themselves not the
spacetime distribution of system of such identical particles.
We will show that in this case the character 
of quantum interaction could be judged in respect 
with the geometry of the state space. This point of view
based on the comprehension that in spite of the Schr\"odinger 
case where any potential forming the metric of the 
configuration space of N identical material points was 
available, at the deeper quantum 
level there is quite definite form of the self-interacting
``prepotential'' which is connected with the geometry of 
single quantum particle state space. Furthermore, if one uses 
the fixed mass of a quantum particle like  the 
phenomenological parameter of some quantum problem
then arises de-Broglie-Schr\"odinger  
wave description in some effective potential.
But if one wishes to understand the mass (inertia)
as a reaction of a quantum object on the deformation
of quantum state \cite{Le1,Le2,Le3}, then one
should take into account {\it entanglement  of
internal degrees of freedom which are the source of 
the inertia (energy, proper mass)} and have to have
self-interaction potential corresponding to this 
entanglement. In general one suspects that {\it there are no
so-called ``elementary particles'' but there the set
of  ``elementary degrees of freedom'' whose entanglement
gives observable quantum numbers and properties
of quantum particles}.

From the technical point of view the new approach to quantum dynamics
is based on understanding that it is impossible to 
build consistent quantum theory basing on spacetime propagation
quantum particles and corresponding waves only. It is the consequence 
of a fact that even most fundamental
notions like ``time-of-arrival'' or ``position-of-event''
which are intuitively
absolutely clear in the frameworks of SR or GR, in quantum area
have very restricted sense in best,
or even do not exist at worst, because it is 
very difficult to endow the notion of ``position-of-arm of clock''
by a sense, if one deals with quantum particles of high energy 
\cite{Aharonov1}. 
Then at the microlevel
all local spacetime structure  is flexible and one suspects
that we need internal classification of quantum motions
(superrelativity in the
class of unitary motions of quantum states which leaves intact the shape 
of ``ellipsoid of polarization'').  Then these transformations take 
the place 
of expended N-level ``spacetime symmetries'' instead of  Lorentz  
symmetries which looks now merely like a ``dipole approximation'' 
in terms of 
2-level system (logical spin =1/2) in 4 dimension classical spacetime. 
In this case only 
{\it quantum transition (deformation of quantum state are represented 
as the deformation of the ``ellipsoid of polarization'')} 
has an objective 
sense of quantum transition which is primary relative to the secondary 
the fact of its registration.  

Then question arises: what we must put in the base of new theory
if, in fact, neither Hamiltonian nor Lagrangian approach are longer
applicable (at least in ordinary sense) to real quantum world?
Our proposal is based on the two postulates:

1. The principle of superrelativity is applicable to quantum 
states in 
respect with a new classification of quantum motions in  complex 
projective space $CP(N)$  \cite{Le1,Le2,Le3}.

2. The action of quantum system comprises two parts in the 
tangent fiber bundle over $CP(N)$: ``vertical'' part connected
to gauge transformation of invariant ellipsoid of polarization,
and ``horizontal'' part, connected to deformation of quantum
states (geometrically it looks like deformation of the
ellipsoid of polarization). This distinction based on invariant 
classification of quantum motions as was mentioned above.  

Hence, I believe that at deep enough quantum level
the problem of dynamical measurement evokes a new quantum 
geometry which is not connected 
directly with spacetime geometry but their relationships 
must lead to observable relativistic force limit in 
ordinary spacetime. This new quantum 
geometry connected with {\it geometry of the symmetry group 
stucture} and geometry of representation space of this
group. 

\section{Absolutely Rigid Scale in Quantum Area}
In order to explain why we should build the quantum
physics on the basis of geometry of quantum states,
I would like to give a simple speculation.

Classical physics and special relativity based on the 
notion of the {\it absolutely rigid scale. It means
that dynamically our scales (rod) are undeformable,
i.e. the energy of its deformation is infinite}. This
infinity of the energy deformation is, of course, merely 
acceptable idealization in the framework of 
classical physics. On the deep quantum level
this idealization is too rough in any sense. Generally we
should remember that the notion of ``absolutely rigid scale'' is 
ambiguous. In special relativity the rigid scale is 
{\it kinematically contractible} in the moving
frame. In general relativity the notion of rigid scale
is not so clear. But much more difficult
situation arises in the quantum area
where even definition of the ``time-of-arrival''
is misty \cite{Aharonov1}. In fact, it has been shown that
dynamical rigid scale in quantum theory is nonsense
because in spite of classical physics {\it the quantum
spacetime scale is energy (action)--dependent}.
It means that the increasing of energy of interaction
leads to the defreezing of hidden degrees of freedom
(strange, color, beauty, etc.). {\bf Therefore one has not 
only the contraction of de Broglie's wave
length but to the increasing dimensionality of a 
configuration space}. In QFT this is well known effect
of creation and annihilation of quantum particles.
Hence the new (relative Poincare-invariant pure electromagnetic
interaction)  kinds of fundamental interactions,
require a new (relative SR of Einstein) agreements
between macroscopic kinematics,
dynamics, and symmetries of these interactions.
{\bf  In order to take into account these effects we 
should use a physically reasonable geometry (measure) 
capable to withstand dramatic evolution of interacting
quantum system}. Therefore, it is natural to 
leave at first step spacetime description and to 
concentrate the attention  on the dynamics 
of quantum degrees of freedom in quantum state
space of nonlinear fundamental unified field.

\section{The Concept of Quantum State Deformation}
I would like use pure geometric method of description
of quantum states. This approach has more or less
simple geometric interpretation and it makes its spirit
very close to the spirit of classical physics.

We will assume that pure quantum state is a ray,
i.e. point of the projective Hilbert space $CP(N)$.
It has been shown that there is a transition dynamics 
which is determined by an hidden quantum ``potential'' 
arising from the geometry of complex projective 
Hilbert space $CP(N)$ of rays with coordinates 
$\pi^i$ \cite{Le1,Le2,Le3}. 
In the framework of this geometry the form of 
excitations is determined by non-effective action
$SU(N+1)$ arises due to
existence so-called {\it isotropy group} 
$H=U(1)\times U(N)$ for every state vector. Of course,
the parameterization of the H-subgroup elements is 
``local'' in the sense of the dependence on the chosen state 
vector.
This leads to the fact that only transformations 
from the coset manifold $G/H=SU(N+1)/S[U(1)\times U(N)]$
act effectively {\bf (as excitations)} on this state.
It is very important to note, that 
$G/H=SU(N+1)/S[U(1)\times U(N)]$ manifold has not
group structure since $H$ is not a normal subgroup
of $G$ (the coset is topological equivalent of 
the complex projective space $CP(N)$).
Therefore the coset determines quite natural mechanism 
of the unitary symmetry breakdown. 
{\bf Here we have the crucial point: instead of quantum 
dynamics in configuration space $R^{3n}$ of material 
points which modeling pointwise electrons etc., one has 
quantum dynamics in the $SU(N+1)$-group generalized 
coherent state submanifold of whole quantum system. 
Then arises self-consistent potential connected with the
geometry of $CP(N)$ and with the structure of generalized 
coherent quantum state}.
Thereby one has very clear classification of the 
$N^2+2N$ generators
of $SU(N+1)$, namely: $N^2$ those that leave
our state intact and $2N$ those that {\it deform this 
state} \cite{Le1,Le2,Le3}.

That is naturally to rely on the internal 
geometry of the state space $CP(N)$ where
non-effectively acts $SU(N+1)$. These fields 
form generalized coherent state with the minimal 
uncertainty. The obvious image for this state and
its deformation is ``ellipsoid of polarization''.
{\bf The parameters of the 
ellipsoid of polarization comprise of state vector variables itself
and the ``orientation'' of the quantum frame relative
a measuring device}.

To summarize we can say that one has {\it the reconstruction}
of the unitary symmetry $SU(N)$ in ``superrelativity'' which
means {\bf the conservation of ellipsoid polarization shape
relative isotropy group of a generalized ``spin coherent state''}.
Such reconstruction of the unitary symmetry represents a concrete
mechanism of symmetry breakdown with help the coset 
transformations   $G/H=SU(N+1)/S[U(1)\times U(N)]$
up to isotropy  group  $H=U(1)\times U(N)$. Therefore {\bf the shape
of the ellipsoid of polarization is a new integral invariant
of unitary quantum dynamics}. We will try to connect this 
this mechanism of the unitary symmetry breakdown
with the problem of mass split effects in unitary fundamental
multiplets of  ``elementary particles''.
 
\section{Dynamical Principles and Dynamical Variables as Vector Fields}
The following main dynamical principle I will put in the base
of quantum theory:

Variational principle of the least action in the 
tangent fiber bundle over complex projective state space 
$CP(N)$. The action contains two components: ``horizontal''
which represented by a geodesic
of Fubini-Study metric in $CP(N)$; and ``vertical'' corresponding
pure gauge of reorientation of the ``ellipsoid of polarization''.
Very important: instead of 
integration in full spacetime one should integrate only 
between two points (initial and final states) along
the lift of geodesic in the tangent fiber bundle (see
for example \cite{Taylor} taking into account that there
base manifold is spacetime but in our case it is
$CP(N)$).

Therefore, dynamical symmetry breakdown realized in deformation
of quantum state may be treated as a 
classical force analog in quantum theory. Just these 
transformations from the coset give a main contribution 
in the action of matter. 

In the framework of the local state-dependent approach
one can formulate a quantum
scheme with help of more flexible mathematical structure than
matrix formalism. I mean matrix elements of transitions 
between {\it two arbitrary far states} are associated with
{\bf in fact bi-local dynamical variables} that
bring a lot of technical problems in quantum field area.
However the 
infinitesimal {\bf local dynamical 
variables} related to deformations of quantum
states are well  defined in projective Hilbert 
space as well as quantum states itself. 
They are local tangent vector fields to the 
projective Hilbert space $CP(N)$ which 
correspond to the group variation of the 
relative ``Fourier components'', i.e. 
{\it generators of group-differential operators 
of first order} \cite{Le1,Le2,Le3}. 

Our generators (q-numbers) realize non-linear representation of 
the unitary group. Here I would like to point out only
one important from the physical point of view link
between curvature of the projective Hilbert space 
$CP(N)$, fundamental, and adjoint  realization of $SU(N+1)$
group as the set of $N^2+2N$ infinitesimal operators.

In the local coordinates $\pi^i = \frac{\Psi^i}{\Psi^0}$
one can build the infinitesimal generators of the
Lie algebra $AlgSU(N+1)$.
They span the tangent Hilbert space.
The coefficients of these tangent vectors are defined 
by the analytic composition function 
$\pi^i=\pi^i(\Psi;\epsilon)$ and by the corresponding
``velocity field''.
But instead of the ``velocity field'' $U_{\sigma}^i(\Psi)$
in homogeneous coordinates $\Psi$ \cite{Gourdin}
we use ``velocity field''
$\Phi_{\sigma}^i(\pi)$ in the non-homogeneous local projective 
coordinates $\pi$. Then one has to use explicit form of 
$\Phi^i_\sigma$ for $N^2+2N$ infinitesimal generators of 
the Lie algebra $AlgSU(N+1)$. For example for the three-level
system, algebra $SU(3)$ has 8 infinitesimal generators which 
are given by:
\begin{eqnarray}
D_1(\lambda)=i \frac{\hbar}{2}[[1-(\pi^1)^2]\frac{\delta}{\delta \pi^1}
-\pi^1 \pi^2 \frac{\delta}{\delta \pi^2}
-[1-(\pi^{1*})^2]\frac{\delta}{\delta \pi^{1*}}
+\pi^{1*} \pi^{2*} \frac{\delta}{\delta \pi^{2*}}] ,  \cr
D_2(\lambda)=- \frac{\hbar}{2}[[1+(\pi^1)^2]\frac{\delta}{\delta \pi^1}
+\pi^1 \pi^2 \frac{\delta}{\delta \pi^2}
+[1+(\pi^{1*})^2]\frac{\delta}{\delta \pi^{1*}}
+\pi^{1*} \pi^{2*} \frac{\delta}{\delta \pi^{2*}}] , \cr
D_3(\lambda)=-\hbar[\pi^1 \frac{\delta}{\delta \pi^1}+\frac{1}{2}\pi^2 \frac{\delta}{\delta \pi^2}
+\pi^{1*} \frac{\delta}{\delta \pi^{1*}}+\frac{1}{2}\pi^{2*} \frac{\delta}{\delta \pi^{2*}}], \cr
D_4(\lambda)= i\frac{\hbar}{2}[[1-(\pi^2)^2]\frac{\delta}{\delta \pi^2}
-\pi^1 \pi^2 \frac{\delta}{\delta \pi^1}
-[1-(\pi^{2*})^2]\frac{\delta}{\delta \pi^{2*}}
+\pi^{1*} \pi^{2*} \frac{\delta}{\delta \pi^{1*}}] , \cr
D_5(\lambda)= -\frac{\hbar}{2}[[1+(\pi^2)^2]\frac{\delta}{\delta \pi^2}
+\pi^1 \pi^2 \frac{\delta}{\delta \pi^1}
+[1+(\pi^{2*})^2]\frac{\delta}{\delta \pi^{2*}}
+\pi^{1*} \pi^{2*} \frac{\delta}{\delta \pi^{1*}}], \cr
D_6(\lambda)=i\frac{\hbar}{2}[\pi^2 \frac{\delta}{\delta \pi^1}
+\pi^1 \frac{\delta}{\delta \pi^2}
-\pi^{2*}\frac{\delta}{\delta \pi^{1*}}
-\pi^{1*} \frac{\delta}{\delta \pi^{2*}}] , \cr
D_7(\lambda)=\frac{\hbar}{2}[\pi^2 \frac{\delta}{\delta \pi^1}
-\pi^1 \frac{\delta}{\delta \pi^2}
+\pi^{2*}\frac{\delta}{\delta \pi^{1*}}
-\pi^{1*} \frac{\delta}{\delta \pi^{2*}}] , \cr
D_8(\lambda)=-\frac{3^{-1/2}}{2}i\hbar[\pi^2 \frac{\delta}{\delta \pi^2}
-\pi^{2*} \frac{\delta}{\delta \pi^{2*}}]. 
\label{D8} 
\end{eqnarray}
This Lie algebra of vector fields paves a way to invariant 
classification of quantum particles  based on coherent 
superposition of {\it quantum integral 
of motions in the state space}.
Such integrals of motions should be identified with isospin, charge,
hypercharge, etc. We will show that during procedure finding 
{\it independent invariants} \cite{Olver}
these vector fields may be simplified. It is very important
that these integrals are expressed in terms of complex 
coordinates of coherent states. 

Lets start with simplest vector field
\begin{equation}
D_8(\lambda)=-\frac{3^{-1/2}}{2}i\hbar[\pi^2 \frac{\delta}{\delta \pi^2}
-\pi^{2*} \frac{\delta}{\delta \pi^{2*}}]. 
\label{D18} 
\end{equation}
Using decomposition   
$\pi^2=x+iy$ and 
$\frac{\delta}{\delta \pi^2}=1/2(\frac{\delta}{\delta x} -
 i\frac{\delta}{\delta y})$,
one expresses this vector field (first term) as follows:
\begin{equation}
\pi^2 \frac{\delta}{\delta \pi^2}=1/2[x \frac{\delta}{\delta x} +
 y \frac{\delta}{\delta y} + i(y \frac{\delta}{\delta x} -
 x \frac{\delta}{\delta y})].
\end{equation}
The local invariant $\omega(\pi)$ of $SU(3)$ group is solution 
of the linear  homogeneous equation in partial derivatives
\begin{equation}
\frac{\delta \omega(\pi)}{\delta \pi^2}=1/2[x \frac{\delta \omega(\pi)}
{\delta x} +
 y \frac{\delta \omega(\pi)}{\delta y} + i(y \frac{\delta \omega(\pi)}
{\delta x} -
 x \frac{\delta \omega(\pi)}{\delta y})]=0.
\end{equation}
The general solution one can find by integration of so-called
system of characterisic equation 
\begin{eqnarray}
\frac{\delta x}{x}= \frac{\delta y}{y} \cr
\frac{\delta x}{y}= \frac{\delta y}{-x}
\end{eqnarray}
Two integrals here are $k=\tan \beta = y/x$ and $r^2=|\pi|^2=x^2 + y^2$.
Now we can rewrite our vector field in terms of these two
integrals:
\begin{equation}
\pi^2 \frac{\delta}{\delta \pi^2}=1/2[x \frac{\delta}{\delta x} +
 y \frac{\delta}{\delta y} + i(y \frac{\delta}{\delta x} -
 x \frac{\delta}{\delta y})]=r \frac{\delta}{\delta r} +
 i(1+k^2) \frac{\delta}{\delta k}. 
\end{equation}
The common integral of this vector field and vector field 
$D_3(\lambda)$ is coherent state itself 
$(\pi^1=|\pi^1| e^{i \alpha}, \pi^2=|\pi^2| e^{i \beta})$ 
since {\it these vector fields are the 
generators of the isotropy group of this coherent state} \cite{Le1}.
Note, it is very convenient and important to know explicit expression
for a coherent state stabilizer because often we deal with
just with superposition state, not stationary configuration
like trivial $(1,0,0), (0,1,0)$ etc. In this particular case one has an 
important result: any holomorphic function of $\pi^1, \pi^2$ is local 
invariant of the isotropy group. Therefore, we can express 
the charge operator $Q$ in terms of out vector fields    
\begin{eqnarray}
Q=-\hbar[\pi^1 \frac{\delta}{\delta \pi^1}+\frac{1}{2}\pi^2 \frac{\delta}{\delta \pi^2}
+\pi^{1*} \frac{\delta}{\delta \pi^{1*}}+\frac{1}{2}\pi^{2*} \frac{\delta}{\delta \pi^{2*}}] \cr
-\frac{3^{-1/2}}{2}i\hbar[\pi^2 \frac{\delta}{\delta \pi^2}
-\pi^{2*} \frac{\delta}{\delta \pi^{2*}}]\cr
=-\hbar[\pi^1 \frac{\delta}{\delta \pi^1}+
(\frac{1}{2}-\frac{3^{-1/2}}{2}i )\pi^2 \frac{\delta}{\delta \pi^2}
+\pi^{1*} \frac{\delta}{\delta \pi^{1*}}+
(\frac{1}{2}+ \frac{3^{-1/2}}{2}i) \pi^{2*} \frac{\delta}
{\delta \pi^{2*}}].
\label{D118} 
\end{eqnarray}
The charge operator has now not only algebraic sense but
dynamical content related to dynamics of the coherent 
superposition of
FF. Now any physically important operator may be expressed
in same manner through $D_\sigma$.
Lets express vector field $D_1(\lambda)$ through two local invariants
$\rho = \sqrt{u^2 + v^2}$ and $l = \tan \alpha =u/v$.
Under denotations $\pi^1=u + i v$ and 
$\frac{\delta}{\delta \pi^1}=1/2(\frac{\delta}{\delta u} -
 i\frac{\delta}{\delta v})$
one has
\begin{eqnarray}
[1 - (\pi^1)^2]\frac{\delta}{\delta \pi^1}=
1/2\{[(1-u^2-v^2)\frac{\delta}{\delta u} - 2uv \frac{\delta}{\delta v}]
-i[(1-u^2-v^2)\frac{\delta}{\delta v} + 2uv \frac{\delta}{\delta u}]\}
 \cr =1/2\{\cos \alpha [1-\rho^2 \cos 2\alpha ]\frac{\delta}
{\delta \rho}
- \rho^{-1} \sin \alpha [1+ \rho^2 \cos 2 \alpha] 
\frac{\delta}{\delta \alpha}\} \cr
-i/2 \{\sin \alpha [1+\rho^2 \cos 2\alpha ]\frac{\delta}{\delta \rho}
+\rho^{-1} \cos \alpha [1- \rho^2 \cos 2 \alpha] 
\frac{\delta}{\delta \alpha}\}. 
\end{eqnarray}
Therefore we have two characteristic equations
\begin{eqnarray}
\frac{d \rho}{\cos \alpha [1-\rho^2 \cos 2\alpha]} =
\frac{- \rho d \alpha}{\sin \alpha [1+ \rho^2 \cos 2 \alpha]} 
\cr
\frac{d \rho}{\sin \alpha [1+\rho^2 \cos 2\alpha]} =
\frac{\rho d \alpha}{\cos \alpha [1-\rho^2 \cos 2 \alpha]} 
\end{eqnarray}
Trying to solve for example first of them by the two steps 
of simplification: 

1) multiplying both parts of the equation 
\begin{equation}
\frac{d \rho}{d \alpha} =-\rho \frac{1}{\tan \alpha}
\frac{1- \rho^2 \cos 2 \alpha} {1+\rho^2 \cos 2\alpha} 
\end{equation}
by  $\rho$ and assuming $\sigma = \rho^2$, one obtain 
\begin{equation}
-2 \sigma \frac{1}{\tan \alpha} [1- \rho^2 \cos 2 \alpha] d \alpha = 
[1+\rho^2 \cos 2\alpha ] d \sigma; 
\end{equation}

2) multiplying both parts by $\sin ^2 \alpha $, one obtains
\begin{equation}
- \sigma d (\cos 2 \alpha )[1- \rho^2 \cos 2 \alpha]= 
\sin ^2 \alpha [1+\rho^2 \cos 2\alpha ] d \sigma; 
\end{equation}
which we can rewrite in more simple form
\begin{equation}
\frac{d \sigma}{d p }=
 -2 \frac {\sigma}{p-1} \frac{p \sigma -1}{p \sigma +1},
\end{equation}
where we put $p = \cos 2 \alpha$,
which looks like not integrable. Probably the reason is that
our vector field is not
defying one-parameter group, but coset transformation
$G/H=SU(N+1)/S[U(1)\times U(N)]$.  
Now we would like to analyze vector field 
\begin{eqnarray}
\pi^1 \pi^2 \frac{\delta f}{\delta \pi^2} = 1/2 (u+iv)(x+iy)
(\frac{\delta f}{\delta x} - 
i \frac{\delta f}{\delta y})\cr
= 1/2 \{[u (x \frac{\delta f}{\delta x} 
+ y \frac{\delta f}{\delta y})
- v (y \frac{\delta f}{\delta x} 
- x \frac{\delta f}{\delta y})] \cr
+i[u (y \frac{\delta f}{\delta x} 
- x \frac{\delta f}{\delta y})
+ v( x \frac{\delta f}{\delta x} 
+ y \frac{\delta f}{\delta y})]\} \cr
= 1/2 \{[u r\frac{\delta f}{\delta r }
+ v (1+ k^2) \frac{\delta f}{\delta k }]
+ i[v r\frac{\delta f}{\delta r }
- u(1+ k^2) \frac{\delta f}{\delta k }].
\end{eqnarray}
That is as before, we have two characteristic equations for 
the real and imaginary parts correspondently  
\begin{eqnarray}
\frac{d r}{u r}= \frac{d k}{ v (1+ k^2)} \cr
\frac{d r}{ v r}= \frac{d k}{ - u (1+ k^2)}.
\end{eqnarray}
Two integrals $I(\alpha, \beta, r)= r e^{\beta \tan \alpha}$ and
$J(\alpha, \beta, r)= r e^{- \frac{\beta}{ \tan \alpha}}$ are
not, however, functionally independent, since we have
the dependence
\begin{equation}
\ln (\frac{I}{r}) \ln (\frac{J}{r}) + \beta^2 = 0.
\end{equation}
Therefore, we can, as before, express 
$\pi^1 \pi^2 \frac{\delta f}{\delta \pi^2}$ in terms of 
the one of the  invariants $I$ or $J$. It is clear this
procedure is available for any dimension. 

Two interesting questions arise:

1. How is possible to build
from the local invariants such global invariant 
as ellipsoid of polarization?

2. May the non-integrability of characteristic equations
to lead to chaotic (stochastic) behavior?

Answer on the first questions is unknown.
Answer on second question should be based on the analysis of
geodesic behavior in $CP(N)$. Therefore, the curvature of the
$CP(N)$ should be taken into account. 
 
\section{The Curvature of CP(N)}
A {\it commutator of two dynamical variables} is one of the 
fundamental notions of quantum theory. Now it arises 
as a {\it commutators of the tangent vector fields
to CP(N)}. But commutator is non-invariant
relative the shifts of the tangent vector fields. Then 
the {\it curvature} of the projective
Hilbert space begins to play a very important role
\cite{Le1,Le2,Le3}. Namely, curvature is  ``history'' invariant  
relative such shifts, i.e. pure local in the state space.
If one treats the differentiation of 
some state-dependent function
$F(\pi^1,...,\pi^N, \pi^{*1},...,\pi^{*N})$
in respect with $\pi^i,\pi^{*i}$ as a {\it variation}
then one should treat the {\it independence of the 
curvature operator}
\begin{equation}
R(X,Y)Z=[\nabla_X,\nabla_Y]Z-\nabla_{[X,Y]}Z
\end{equation}
{\it as a pre- and post- history independent operator due to the compensation term}  $\nabla_{[X,Y]}$ {\it in some particular
coherent state} $(\pi^1_0,...,\pi^N_0, \pi^{*1}_0,...,\pi^{*N}_0)$ {\it relative to
the shift transformations} 
\begin{eqnarray}
X' = X + A(\pi^1,...,\pi^N, \pi^{*1},...,\pi^{*N}),\cr
Y' = Y + B(\pi^1,...,\pi^N, \pi^{*1},...,\pi^{*N}),\cr
Z' = Z + C(\pi^1,...,\pi^N, \pi^{*1},...,\pi^{*N}),
\end{eqnarray}
where $A=a^i (\pi^1,...,\pi^N, \pi^{*1},...,\pi^{*N})  \frac{\partial}{\partial \pi^i}
+  a^{*i} (\pi^1,...,\pi^N, \pi^{*1},...,\pi^{*N})\frac{\partial}{\partial \pi^{*i}} $
and $a^i (\pi^1,...,\pi^N, \pi^{*1},...,\pi^{*N}) $ are smooth arbitrary functions 
which are zero in the mentioned above particular coherent state
$a^i (\pi^1_0,...,\pi^N_0, \pi^{*1}_0,...,\pi^{*N}_0)=0 $.
This is a very important property of the new dynamics because
vector fields representing dynamical variables may be shifted during
some interactions in previous history. For us, however, only invariant
characteristics of interaction are interesting. The curvature supplies
us just by the invariant mechanism of interaction.

In order to see the explicit dependence of the spectrum
of the operator of the curvature
\begin{equation}
R(X,Y)=[\nabla_X,\nabla_Y]-\nabla_{[X,Y]}
\label{curvoper}.
\end{equation}
one should take into account the functional dependence
of vector fields $X,Y$ from the local coordinate
because the local unitary parallel transported frame
is {\it anholonomich}. Therefore the
tensor of curvature in the anholonomic frame has 
additional terms. For us will be interesting the 
curvature operator
in 2-dimension direction defined by the local
vector fields (generators) of the projective 
representation of $SU(N+1)$ \cite{Le1,Le2,Le3}. 
It seems to be important just these vector fields
are true {\it physical discriminators of the motions
in} $CP(N)$ because they define the curvature  
\begin{eqnarray}
R(D_\alpha,D_\beta})D_\gamma = 
\{[\nabla_{\Phi^i_\alpha \frac{\partial}{\partial \pi^i}
+ \Phi^{*i}_\alpha \frac{\partial}{\partial \pi^{*i}}},
\nabla_{\Phi^k_\beta \frac{\partial}{\partial \pi^k} 
+ \Phi^{*k}_\beta \frac{\partial}{\partial \pi^{*k}}]\cr
-\nabla_{[\Phi^i_\alpha \frac{\partial}{\partial \pi^i}
+ \Phi^{*i}_\alpha \frac{\partial}{\partial \pi^{*i}},
\Phi^k_\beta \frac{\partial}{\partial \pi^k}
+ \Phi^{*k}_\beta \frac{\partial}{\partial \pi^{*k}}]}\}
D_\gamma
\end{eqnarray}
in $\alpha, \beta$ directions of isotopic space. Here 
$\hat{D}_\sigma$ is
one of the $N^2+2N$ directions in the $SU(N+1)$
group manifold. Then one has
\begin{equation}
D_\sigma =\Phi^i_\sigma (\pi,\pi^*)\frac{\partial}{\partial \pi^i}
+\Phi^{i*}_\sigma (\pi,\pi^*)\frac{\partial}{\partial \pi^{i*}},
\label{dif}
\end{equation}
where
\begin{equation}
\Phi_{\sigma}^i = \lim_{\epsilon \to 0} \epsilon^{-1}
\biggl\{\frac{[\exp(i\epsilon \lambda_{\sigma})]_m^i \Psi^m}{[\exp(i \epsilon \lambda_{\sigma})]_m^j
\Psi^m }-\frac{\Psi^i}{\Psi^j} \biggr\}=
\lim_{\epsilon \to 0} \epsilon^{-1} \{ \pi^i(\epsilon \lambda_{\sigma}) -\pi^i \},
\end{equation}
are the local (in CP(N)) state-dependent components  of the SU(N+1) 
group generators $\lambda_{\sigma}$,
$\epsilon = \frac{E t}{\hbar}$ is dimensionless action
parameter, and $j\neq i$ is number of the local $U_j$-chart 
which are studied in \cite{Le1,Le2,Le3}.

In accordance with the linear properties of the covariant
derivatives and assuming that $X^k(\pi,\pi^*)$ is
complex analytic one has
\begin{eqnarray}
\nabla_{D_\alpha}X^k = 
\nabla_{\Phi^i_\alpha \frac{\partial}{\partial \pi^i}
+ \Phi^{*i}_\alpha \frac{\partial}{\partial \pi^{*i}}}X^k = \Phi^i_\alpha (\frac{\partial X^k}{\partial \pi^i}+
\Gamma^k_{in}X^n)+ \Phi^{*i}_\alpha \frac{\partial X^k}{\partial \pi^{*i}}
\end{eqnarray}
and
\begin{eqnarray}
\nabla_{D_\alpha}X^{*k} = 
\nabla_{\Phi^i_\alpha \frac{\partial}{\partial \pi^i}
+ \Phi^{*i}_\alpha \frac{\partial}{\partial \pi^{*i}}}X^{*k}
= \Phi^i_\alpha \frac{\partial X^{*k}}{\partial \pi^i}+
 \Phi^{*i}_\alpha (\frac{\partial X^{*k}}{\partial \pi^{*i}} + \Gamma^{*k}_{*i*n}X^{*n}).
\end{eqnarray}
For the compensation term one has
\begin{eqnarray}
\nabla_{[\Phi^i_\alpha \frac{\partial}{\partial \pi^i}
+ \Phi^{*i}_\alpha \frac{\partial}{\partial \pi^{*i}},
\Phi^k_\beta \frac{\partial}{\partial \pi^k}
+ \Phi^{*k}_\beta \frac{\partial}{\partial \pi^{*k}}]}X^k 
= [D_\alpha,D_\beta] X^k + C^\gamma_{\alpha,\beta} 
\Phi^i_\gamma \Gamma^k_{in} X^n.
\end{eqnarray}
and
\begin{eqnarray}
\nabla_{[\Phi^i_\alpha \frac{\partial}{\partial \pi^i}
+ \Phi^{*i}_\alpha \frac{\partial}{\partial \pi^{*i}},
\Phi^k_\beta \frac{\partial}{\partial \pi^k}
+ \Phi^{*k}_\beta \frac{\partial}{\partial \pi^{*k}}]}X^{*k} = [D_\alpha,D_\beta] X^{*k} + C^{*\gamma}_{\alpha,\beta} 
\Phi^{*i}_\gamma \Gamma^{*k}_{*m*n} X^{*n}.
\end{eqnarray}
The expression for the tensor of curvature in $CP(N)$
in 2-dimension direction $(\alpha,\beta)$ defined by physical fields
is as follows:
\begin{eqnarray}
R(D_\alpha,D_\beta)X^k = 
[\nabla_{D_\alpha},\nabla_{D_\beta}] X^k -
\nabla_{[D_\alpha,D_\beta]} X^k \cr
=\{(D_\alpha \Phi^i_\beta)\Gamma^k_{in} +
\Phi^i_\beta (D_\alpha \Gamma^k_{in})-
(D_\beta \Phi^m_\alpha)\Gamma^k_{mn}-
\Phi^m_\alpha (D_\beta \Gamma^k_{mn}) \cr +
\Phi^m_\alpha \Gamma ^k_{mp}\Phi^i_\beta \Gamma ^p_{in}-
\Phi^i_\beta \Gamma ^k_{ip}\Phi^m_\alpha \Gamma ^p_{mn}
-
C^\gamma_{\alpha \beta} \Phi^i_\gamma \Gamma^k_{in}\}X^n \cr
=\{(D_\alpha \Phi^i_\beta - D_\beta \Phi^i_\alpha)
\Gamma^k_{in} +
\Phi^i_\beta (D_\alpha \Gamma^k_{in})- 
\Phi^m_\alpha (D_\beta \Gamma^k_{mn}) \cr +
\Phi^m_\alpha \Gamma ^k_{mp}\Phi^i_\beta \Gamma ^p_{in}-
\Phi^i_\beta \Gamma ^k_{ip}\Phi^m_\alpha \Gamma ^p_{mn}
-C^\gamma_{\alpha \beta} \Phi^i_\gamma \Gamma^k_{in}\}X^n
\cr = \{(D_\alpha \Phi^i_\beta - D_\beta \Phi^i_\alpha )
\Gamma^k_{in}  
-(\Phi^i_\beta \Phi^{*s}_\alpha -
\Phi^i_\alpha \Phi^{*s}_\beta)\frac{\partial \Gamma^k_{in}}{\partial \pi^{*s}} \cr +
\Phi^m_\alpha \Gamma ^k_{mp}\Phi^i_\beta \Gamma ^p_{in}-
\Phi^i_\beta \Gamma ^k_{ip}\Phi^m_\alpha \Gamma ^p_{mn}
- C^\gamma_{\alpha \beta} \Phi^i_\gamma \Gamma^k_{in}\}X^n 
\cr =
\{(D_\alpha \Phi^i_\beta - D_\beta \Phi^i_\alpha )
\Gamma^k_{in}  
-(\Phi^i_\beta \Phi^{*s}_\alpha -
\Phi^i_\alpha \Phi^{*s}_\beta) R^k_{i*sn} \cr +
\Phi^m_\alpha \Gamma ^k_{mp}\Phi^i_\beta \Gamma ^p_{in}-
\Phi^i_\beta \Gamma ^k_{ip}\Phi^m_\alpha \Gamma ^p_{mn}
- C^\gamma_{\alpha \beta} \Phi^i_\gamma \Gamma^k_{in}\}X^n, 
\end{eqnarray}
since $R^k_{isn}=\frac{\partial \Gamma^k_{in}}{\partial \pi^s}=0$.
Here $C^\gamma_{\alpha \beta}$ are of $SU(N+1)$ group structure constants.

It has been shown  \cite{Le1,Le2,Le3} that this geometry connected
with the natural classification of motions
which give the possibility to avoid, in fact,
artificial separation quantum system on ``heavy''
and ``light'' subsystems using  dynamical group
$G=SU(N+1)$ and its
breakdown to the isotropy group $H=U(1) \times U(N)$
of the pure quantum state. Besides that, the quantum coordinates 
of the $N+1$-level system $\pi^i$ and 
tangent vector  fields of Goldstone and Higgs 
excitations  $\Phi^i_\sigma (\pi)$ give a full
description of the quantum system itself in the tangent fiber bundle over $CP(N)$ and one does not need any ``second'' quantum system  as a 
reference frame.
Thereby the local movable frame a l\'a Cartan naturally arises in the 
projective Hilbert space and  their coefficients depends only on local projective coordinates $\pi^i$.

\section{Geometrical Bosons and Real Fields}
In our description the real deviation from the 
vacuum state of FF related to rate of the changing of local
vacuum (tangent vector fields). These deviation
(deformation) one can identify with the dynamical variables
of some quantum system. In order to do it,
one have to have a map between underlying dynamical structure
in the tangent fiber bundle over $CP(N)$ and real spacetime
propagation of observable particles and fields.
I emphasized already \cite{Le1} that my  
introduction of the the energy variation which associated with 
infinitesimal gauge transformation of the local 
frame  with the connection
\begin{eqnarray}
\delta H= \frac{1}{\mu}\delta U=\frac{1}{\mu} A_m \delta \pi^m 
=\frac{1}{\mu}\frac{\delta U}{\delta \pi^m}\delta \pi^m
=-\frac{\hbar}{\mu} \Gamma^i_{km}\xi^k \frac{\partial \Psi^a}{\partial \pi^i}\delta \pi^m |a>.
\label{Am} 
\end{eqnarray}
for the droplet model of extended quantum particles 
in  Minkowski spacetime is inconsistent since it was not based 
on the dynamical method of spacetime metrization. 
Dynamical method should include transformation of internal
degrees of freedom into the energy-momentum of a system without
a priori fixed spacetime structure. 

First step in this direction can be done by 
``geometrical bosons'' introduction, taking into account 
\begin{equation}
[\frac{\partial}{\partial \pi^k},\pi^i]_-=\delta^i_k.
\label{comm}
\end{equation}
In order to agree the standard Fock representation
and the definition of vacuum state by a holomorphic 
function $F_{vac}$ we could introduce the simplest ``Hamiltonian''
of {\it geometrical bosons} as the tangent vector fields
\begin{equation}
\Xi^i_k(bos)=\hbar \omega \pi^{*i}\frac{\partial}{\partial \pi^{*k}}.
\label{bos}
\end{equation}
Since $\frac{\partial F_{vac}}{\partial \pi^{*k}}=0$
one has:
\begin{eqnarray}
\Xi^i_k(bos)F_{vac}=0; \cr
\Xi^i_k(bos)\pi^{*s} F_{vac}=\hbar \omega 
\pi^{*i}\delta^s_k F_{vac}; \cr
\Xi^i_k(bos)\pi^{*s_1} \pi^{*s_2}F_{vac}=\hbar \omega \pi^{*i}(\pi^{*s_2}\delta^{s_1}_k +
\pi^{*s_1}\delta^{s_2}_k)F_{vac}; \cr
. \cr
. \cr
. 
\label{excit}
\end{eqnarray}
One can introduce the function of excitations
of different degrees of freedom 
$F(s_1,...,s_{\cal{N}})=\pi^{*s_1} \pi^{*s_2}...
\pi^{*s_{\cal{N}}}F_{vac}$ and the function of a 
multifold excited degree of freedom
$F(s;\cal{N})$ $=(\pi^{*s})^{\cal{N}}F_{vac}$.
It is easy to see that only in very particular 
case we have the coincidence with the equidistant
spectrum of harmonic oscillator. These are
\begin{eqnarray}
\Xi^i_k(bos)F(s;1)=\hbar \omega F(s;1); \cr
\Xi^i_k(bos)F(s;2)=2\hbar \omega F(s;2); \cr
. \cr
. \cr
. \cr
\Xi^i_k(bos)F(s;{\cal{N}})
={\cal{N}} \hbar \omega F(s;{\cal{N}}).
\label{oscill}
\end{eqnarray}
Therefore one can think
that besides ordinary bosons (e.g. photons)
the model of the ``geometrical bosons'' contains
the different kinds of excitations with a
non-equidistant spectrum. This is the 
consequence of the multiplete (anisotropic)
interaction structure of the internal degrees of freedom. 

If one assumes that monochromatic waves associated with
free particles (in de Broglie spirit) then arise a possibility
to reduce locally in $CP(N)$ an arbitrary tangent vector field
$X^i=X^{\sigma}\Phi^i_{\sigma}$ to the vector field \ref{bos}.
Thereby one has the transformation of FF degrees of freedom
into the Hamiltonian of the ``geometric bosons''. 
This is some equivalent of the Hamiltonian 
diagonalization (u-v Bogolubov transformation)  
in ordinary quantum formalism. 
The frequency
takes the place of the ``rate'' of an action variation which is
equivalent to the energy. Momentum may be introduced in the same
manner as a ``ramp'' of an action variation but one  
should know the dispersion law of these geometrical bosons. 
This law should inherit the geometry of the base manifold
$CP(N)$ since these gradients may be measured of the
``vertical'' component of action in fibers. Hence, it may be 
found as a compensation of
our forcible ``horizontal'' transformation of FF degrees of freedom
into the Hamiltonian of the free ``geometric bosons''.
Horizontal transformation is the successive family of ``rotation 
about commutator of the two vector fields'', i.e. about 
Lie derivative transforming arbitrary path in $CP(N)$
into geodesic \cite{Le3}. Locally it may be described by 
the curvature form in $CP(N)$
\begin{eqnarray}
\omega^i_k= \Gamma^i_{km}\delta \pi^m. 
\label{omega} 
\end{eqnarray}
The transformation law in a tangent
Hilbert space
\begin{eqnarray}
\xi^{'i}=U^i_m \xi^m
\label{transXi} 
\end{eqnarray}
leads to the transformations law of the curvature form of the
well know kind:
\begin{eqnarray}
\omega^{'i}_k=  U^{-1i}_m \omega^{m}_n U^n_k
+\delta U^{-1i}_t U^t_k.
\label{transOm} 
\end{eqnarray}
This local (in $CP(N)$) non-Abelian gauge field looks 
like Wilczek--Zee
gauge potential \cite{WZ} but it has, of course, 
different physical sense. The physical status of 
this field is the subject of our interest.
Dynamical description of this gauge
field requires the dynamical introduction of 
spacetime coordinates in the fiber of the tangent fiber
bundle over $CP(N)$.
It is possible because any geodesic on the $CP(N)$
lies in some $CP(1)$ which, therefore, may be treated as
space of coherent states of the ``logical spin'' 1/2 in the
basis of $|1>=yes$, $|0>=no$ kinds of measurements.

\vskip 1cm
ACKNOWLEDGEMENTS

I am sincerely grateful to Larry Horwitz for stimulating interest
to this research.

\end{document}